\documentclass[twocolumn,showpacs,preprintnumbers,superscriptaddress,amsmath,amssymb,aps,prl]{revtex4}

\usepackage{graphicx}
\usepackage{dcolumn}
\usepackage{bm}

\begin{document}

\title{Optimal superluminal systems}

\author{Bruno Macke}
\affiliation{Laboratoire de Physique des Lasers, Atomes et Mol\'{e}cules, CERLA, Universit\'{e} de Lille I, 59655 Villeneuve d'Ascq, France}
\author{Bernard S\'{e}gard }\email {bernard.segard@univ-lille1.fr}
\affiliation{Laboratoire de Physique des Lasers, Atomes et Mol\'{e}cules, CERLA, Universit\'{e} de Lille I, 59655 Villeneuve d'Ascq, France}
\author{Franck Wielonsky}
\affiliation{Laboratoire Paul Painlev\'{e}, B\^{a}timent M2, Universit\'{e} de Lille I, 59655 Villeneuve d'Ascq, France}
\date{\today}
\begin{abstract}
We demonstrate that significant effects in the "superluminal propagation" 
of light-pulses cannot be observed without involving systems whose gain 
explodes outside the pulse spectrum. We explicitly determine the minimum 
norm of the gain to attain given superluminal effects and the transfer function of 
the corresponding optimal system. The gain-norms which would be required with the \emph {most 
efficient} systems considered up to now (dispersive media, photonic barriers) to attain 
the same effects are shown to exceed the minimum by several orders of magnitude. 
We finally estimate the largest superluminal advances which could be attained in a realistic experiment.

\end{abstract}
\maketitle

The apparently superluminal propagation of light-pulses has been observed
with various systems, mainly with systems involving anomalously dispersive
media or photonic barriers. For reviews, see, e.g., \cite{key-1,key-2,key-3,key-4,key-5}.
In these experiments, the envelope of the pulse having covered some
distance $L$ is nearly identical to that of the incident pulse and
in advance of that of a pulse which has covered the same distance
$L$ at the velocity $c$ of light in vacuum. This surprising behaviour
is not at odds with the relativistic  causality. Indeed the signal received at some time $t$ is not the
consequence of the signal emitted at a well-defined time but of all
the signals anterior to $t$ by more than $L/c$. Otherwise
said, there is no cause-to-effect relation between the homologous points
of the envelopes of the incident and transmitted pulses and the widespread
statement that \emph {the pulse maximum leaves the system before it even enters
it} is somewhat misleading. The phenomenon is however quite puzzling
and keeps the subject of an intense theoretical and experimental activity.

In fact Mother Nature resists to a violation of her principles even
when this violation is only apparent and convincing experiments of
superluminal transmission are very difficult to achieve. By convincing
experiments, we mean experiments where (i) the envelopes or the intensity
 profiles of the pulses are detected in real-time and
true-shape (ii) the transmitted pulse is directly compared to the
pulse having propagated at the velocity $c$ (iii) the superluminal
advance $a$ is large compared to the optical period (iv) it is significant
with respect to the pulse duration, say larger than 10\% of the full
width at half maximum (FWHM) of the intensity profile $\tau_{p}$ (v) the
pulse distortion (including noise and parasitic signals) is small
compared to the relative advance $a/\tau_{p}$. Note that (iii) is
a consequence of (i) since the real-time detection of the envelope
requires a time-constant large compared to the optical period. There
are few experiments meeting, even approximately, the previous conditions
\cite{key-6,key-7,key-8,key-9,key-10,key-11,key-12,key-13}. Though
all-optical experiments are possible, only hybrid systems have been
used up to now. They combine an optical part, responsible for the
superluminal effects, and a wide-band electronic device whose function
is to normalise the amplitude of the transmitted pulse. In most experiments,
the transmission of the optical part, usually a resonantly absorbing
medium \cite{key-6,key-9,key-10,key-11,key-12} or a photonic barrier
\cite{key-7,key-8,key-14}, is low and the electronic device is an
amplifier. To our knowledge, only one experiment \cite{key-13}
has evidenced significant superluminal effects with an active optical part
(amplifying medium). The normalisation is then achieved by a suitable
attenuation. In the following, we naturally include the normalisation
device (amplifier or attenuator) in the system under consideration. 

As already noted in previous papers dealing with particular arrangements 
(see, e.g., \cite{key-5}), large superluminal effects are only attained with 
systems whose gain explodes outside the pulse spectrum. We will show that 
this is true for any physically realisable system and determine the lower 
limit to the gain norm required to observe given superluminal effects. This result 
is of special importance since in a real experiment the gain-norm should be limited 
to avoid problems of noise (no matter its origin), of instability and of hypersensitivity 
to parasitic signal and to localised defects in the incident pulse profile \cite{key-6}. 
Conversely restricting the gain to realistic values determines the upper limit 
to the actually observable effects.

The problem is studied in the frame of the linear systems theory \cite{key-15}.
We denote by $e(t)$ and $s(t)$ the envelopes of the incident and
transmitted pulses and by $E(\omega)=\int_{-\infty}^{\infty}e(t)\exp(-i\omega t)dt$
and $S(\omega)$ their Fourier transforms. The envelopes are assumed
to be slowly varying at the scale of the optical period. Their Fourier
transforms are then concentrated around $0$ in a region of width
small compared to the optical frequency. In all the sequel, $t$ designates
the local time, equal to the real time in $e(t)$ and retarded by
the luminal transit time $L/c$ in $s(t)$. The system is
characterised by its impulse response $h(t)$ or its transfer function
$H(\omega)$, such that $s(t)=h(t)\otimes e(t)$ and $S(\omega)=H(\omega)E(\omega)$.
We assume that $E(\omega)$ and $H(\omega)$ have a finite energy
and that $H(\omega)$, Fourier transform of $h(t)$, has a continuation $H(z)$ in the 
complex plane ($z=x+iy=\rho e^{i\theta}$). In our local time picture, the
relativistic causality imposes that $h(t<0)=0$.
Otherwise said, $H(z)$ belongs to $L^{2}(\mathbb{R})$, the Hilbert
space of functions $F(z)$ square summable on the real line $\mathbb{R}$
endowed with the norm $\left\Vert F\right\Vert _{\mathbb{R}}$ such
that $\left\Vert F\right\Vert _{\mathbb{R}}^{2}=\int_{-\infty}^{\infty}\left|F(\omega)\right|^{2}d\omega$
and, more precisely, to the Hardy space $H^{2}(\Pi_{-})$ of functions
$F$ analytic in the lower half-plane $\Pi_{-}$ ($y<0$) which are
Fourier transform of some causal function $f\in L^{2}(0,\infty)$ \cite{key-16}.

We want $s(t)$ to be as close as possible to $e(t+a)$ where $a$
is the superluminal advance ($a>0$). In $L^{2}$ norm, the distortion
is defined by
\begin{equation}
D=\frac{\left\Vert e(t+a)-s(t)\right\Vert \mathbb{_{R}}}{\left\Vert e(t)\right\Vert _{\mathbb{R}}}=
\frac{\left\Vert (H_{a}-H)E\right\Vert _{\mathbb{R}}}{\left\Vert E\right\Vert _{\mathbb{R}}}\label{eq:1}
\end{equation}
where $H_{a}=e^{i\omega a}$ is the transfer function of the non
causal system perfectly achieving the advance $a$ without any distortion.
With a real (causal) system, the distortion will be low if $H(\omega)\approx H_{a}(\omega)$
in the region around $\omega=0$ where $\left|E(\omega)\right|$ is
concentrated. 

To keep tractable calculations, we consider the case $E(\omega)=E_{0}$
for $\left|\omega\right|<\omega_{c}$ and $0$ elsewhere. By taking
$E_{0}=\pi$ and $\omega_{c}=1$, this amounts to take as reference
a pulse of intensity profile $\left|e(t)\right|^{2}=\sin^{2}t/t^{2}$
(FWHM $\tau_{p}=2.78$). The distortion then reads $D=\left\Vert H_{a}-H\right\Vert _{I}/\sqrt{2}$
where $\left\Vert F\right\Vert _{I}$ denotes the norm $L^{2}$ of
$F$ restricted to $I=\left[-1,1\right]$. In the situations of physical
interest $D\ll1$ and $\left\Vert H\right\Vert _{\mathbb{R}}^{2}=
\left\Vert H\right\Vert _{I}^{2}+\left\Vert H\right\Vert _{J}^{2}\approx2+\left\Vert H\right\Vert _{J}^{2}$
where $J=\left[-\infty,-1\right]\cup\left[1,\infty\right]$. In this
model, the problem may then be stated : \emph {given $a>0$ and $D>0$, minimise 
$Q=\left\Vert H\right\Vert _{J}$ under the constraints $H\in H^{2}\left(\Pi_{-}\right)$
and $\left\Vert H _{a}-H\right\Vert _{I}\leq D \sqrt{2}$}.

Based upon a conformal map that sends the unit disk $\mathbb{D}$
$\left(\rho=1\right)$ onto the lower half-plane, we introduce the
map $\widetilde{F}=\Psi\left(F\right)$ defined by
\begin{equation}
\widetilde{F}(z)=\Psi\left(F\right)(z)=\frac{\sqrt{2\pi}}{1-z}F\left(\frac{1+z}{2i(1-z)}\right)\label{eq:2}
\end{equation}
 It is an isometry from $L^{2}(\mathbb{R})$ to the Hilbert space
$L^{2}(\mathbb{T})$ of the unit circle $\mathbb{T}$ endowed with
the norm $\left\Vert F\right\Vert _{\mathbb{T}}$ such that $\left\Vert F\right\Vert _{\mathbb{T}}^{2}=
\int_{0}^{2\pi}\left|F(e^{i\theta})\right|^{2}d\theta/2\pi$.
It sends the subspace $H^{2}(\Pi_{-})$ onto the corresponding Hardy
space $H^{2}(\mathbb{D})$ of the unit disk $\mathbb{D}$ . We denote
by $\widetilde{I}$ and $\widetilde{J}$ the subarcs of $\mathbb{T}$, transforms
of $I$ and $J$ by the map $\Psi$. Then this map allows one to restate
the problem in the unit disk $\mathbb{D}$ instead of the lower half-plane
: \emph {given $a>0$ and $D>0$, minimise $Q=\left\Vert \widetilde{H} \right\Vert _{\widetilde{J}}$
under the constraints $\widetilde{H}\in H^{2}\left(\mathbb{D}\right)$
and $\left\Vert \widetilde{H}_{a}-\widetilde{H}\right\Vert _{\widetilde{I}}\leq D \sqrt{2}$}.

Stated with a general function $\widetilde{K}\in L^{2}(\widetilde{I})$ instead
of the particular $\widetilde{H}_{a}$, this question has been originally
considered in \cite{key-16} and more recently in \cite{key-17},
with important extensions. The solution $\widetilde{H}_{opt}$ of the problem exists and is unique.
Note that, in our case ($\widetilde{K}=\widetilde{H}_{a}$), the constraint 
$\left\Vert H _{a}-H\right\Vert _{I}\leq D \sqrt{2}$ is saturated,
i.e. $\left\Vert H _{a}-H\right\Vert _{I}= D \sqrt{2}$. The solution
$\widetilde{H}_{opt}$ can formally be written under the analytic form
\cite{key-17}:
\begin{equation}
\widetilde{H}_{opt}=\left(1+\lambda\Phi\right)^{-1}P_{H^{2}}(\hat{H}_{a})\label{eq:3}
\end{equation}
 In this expression $\hat{H}_{a}$ is defined as $\widetilde{H}_{a}$
on $\widetilde{I}$ and 0 on $\widetilde{J}$, $P_{H^{2}}$ denotes the orthogonal
projection from $L^{2}(\mathbb{T})$ onto $H^{2}(\mathbb{D})$ and
$\Phi$ is the so-called Toeplitz operator \cite{key-17} acting
on $H^{2}(\mathbb{D})$. It is such that $\Phi(\widetilde{F})=P_{H^{2}}(\check{F})$
where $\check{F}$ is defined as $\widetilde{F}$ on $\widetilde{J}$ and
0 on $\widetilde{I}$. Finally $\lambda\in\left[-1,\infty\right]$ is
an implicit parameter. It is the unique real number such that $\left\Vert H _{a}-H\right\Vert _{I}= D \sqrt{2}$.

From a computational viewpoint, it appears natural to consider $Q$
and $D$ as functions of $\lambda$ \cite{key-17}. It follows
from Eq.\ref{eq:3} that $Q$ and $D$ respectively increases and decreases as $\lambda$ decreases.
As $\lambda\rightarrow-1$, $Q\rightarrow\infty$ while $D\rightarrow0$.
In physical terms, this confirms that a low distortion will always
be paid at the price of a large gain-norm. We have then $\left\Vert H_{opt}\right\Vert _{\mathbb{R}}=
\left\Vert \widetilde{H}_{opt}\right\Vert _{\mathbb{T}}\approx Q$. 

Given $a$ and $D$, the previous analysis leads to the following algorithm for the computation
of the minimum gain norm $Q$ and the corresponding function $\widetilde{H}_{opt}$ :
(i) Choose $\lambda <-1$ and compute $\widetilde{H}_{opt}$ given by Eq.\ref{eq:3} 
(ii) Compute $D$. If it is too large (resp. small), decrease (resp.
increase) $\lambda$. Go to (i). Such a dichotomy algorithm has been
implemented in the software package \emph{Hyperion} developed at INRIA (Institut
National de Recherche en Informatique et Automatique) by the APICS
team \cite{key-18}. See also \cite{key-19} for a closely related algorithm. Eq.\ref{eq:3}, which is infinite
dimensional, is approached by truncating the expansions of the involved
functions so as to consider only their Fourier coefficients of indices
$-N\leq j\leq N$. The optimal transfer function $H_{opt}(\omega)$ is finally obtained by inverting
Eq.\ref{eq:2}: 
\begin{equation}
H_{opt}(\omega)=\frac{\sqrt{2/\pi}}{2i\omega+1}\widetilde{H}_{opt}\left(\frac{2i\omega-1}{2i\omega+1}\right)\label{eq:4}
\end{equation}
Note that $H_{opt}(\omega)$ behaves as $1/i\omega$ for
$\left|\omega\right|\rightarrow\infty$. This behaviour is that of
a first order filter as used in every detection chain. Any further filtering
of the high frequencies will obviously damage the performances of
the system. 
To close this short presentation of our minimisation procedure, 
we remark that it mainly lies on the separation of the spectral domains where the distortion and the gain-norm 
are computed. We have chosen the pulse profile leading to the simplest calculations but the 
procedure might be adapted to any pulse provided that its Fourier transform has a compact support.
\begin{figure}[htbp]
  \begin{center}
    \includegraphics[angle=-90,width=8.5cm]{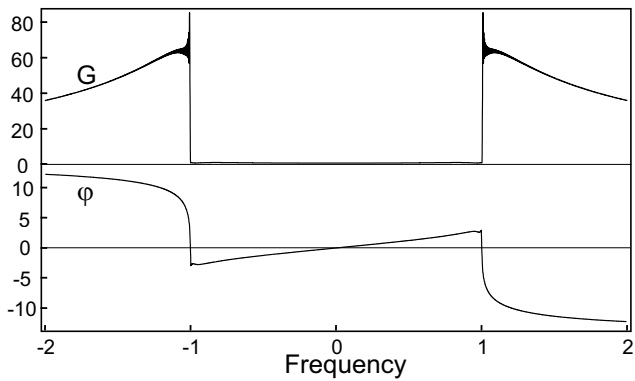}
    \caption{Amplitude-gain $G$ and phase $\varphi$ (radian) of the optimal system as functions of the frequency. 
Parameters: $a=\tau_{p}$ and $D=15\%$. \label{fig1}}
\end{center}
\end{figure}

\begin{figure}[htbp]
  \begin{center}
    \includegraphics[angle=-90,width=8.5cm]{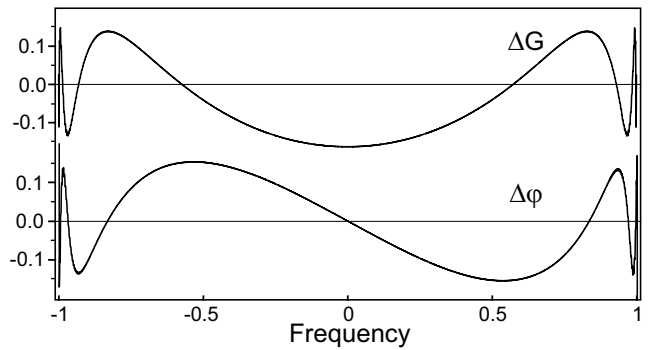}
    \caption{Frequency-dependence of  $\Delta G=G-1$ and of $\Delta\varphi=\varphi-\omega a$ in the useful band. The group advance $a_{g}$
deviates from $a$ by $\Delta a=d(\Delta\varphi)/d\omega\mid{}_{\omega=0}$, that is $\Delta a\approx-0.40$ and $\Delta a/a\approx-14\%$. 
Parameters as in Fig.\ref{fig1}.\label{fig2}}
\end{center}
\end{figure}

\begin{figure}[htbp]
  \begin{center}
    \includegraphics[angle=-90,width=8.5cm]{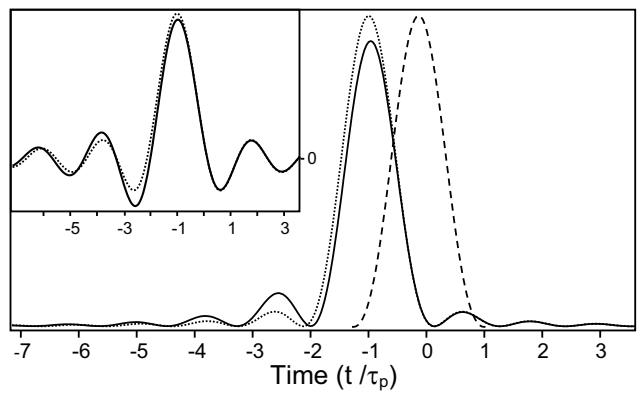}
    \caption{Intensity profile of the pulse transmitted by the optimal system (full line). 
The profiles of the incident pulse advanced by $a=\tau_{p}$ (dotted line) and the main lobe of the incident pulse (dashed line)  are given 
for reference. Insert: Envelopes $s(t)$ (full line) and $e(t+a)$ (dotted line). Parameters as in Fig.\ref{fig1}.\label{fig3}}
\end{center}
\end{figure}

Calculations of the minimum gain-norm $Q$, of the corresponding
transfer function $H_{opt}(\omega)$ and of the transmitted signal
$s(t)$ have been made for $a/\tau_{p}$ (resp. $D$) ranging from
$0.36$ to $2.2$ (resp. $2$ to $30$\%). Satisfactorily enough, the optimal 
system would allow one to conciliate significant advance, moderate distortion and 
reasonable gain. For instance $a=\tau_{p}$ with $D=15$\% would be obtained for $Q=100$. 
Fig.\ref{fig1} shows the overall frequency-dependence of the amplitude-gain $G\left(\omega\right)=\left|H_{opt}(\omega)\right|$
and of the phase $\varphi\left(\omega\right)=\arg\left[H_{opt}(\omega)\right]$ in this reference case. 
As expected, the gain reaches its peak-value near the frontiers of the "stop band" 
(in fact the useful band for superluminal systems). The short ringing close to these frontiers originates from the 
finite number of Fourier coefficients used in the calculations ($N=2000$). The asymptotic values of 
the phase are $\varphi=\pm 9\pi/2$ for $\omega=\mp\infty$, in agreement with Eq.\ref{eq:4}. 
The extra phase-rotation of $8\pi$ entails that $H_{opt}(z)$ has four zeros in the half-plane $y<0$ and, 
consequently, that $H_{opt}(\omega)$ is not minimum-phase \cite{key-15}. The differences $\Delta G=G-1$ 
and $\Delta\varphi=\varphi-\omega a$ for $-1<\omega<1$ (Fig.\ref{fig2}) illustrate how $H_{opt}(\omega)$ deviates
from the ideal transfer function $H_{a}=e^{i\omega a}$ in the useful band. We remark that the group 
advance $a_{g}=d\varphi/d\omega\mid{}_{\omega=0}$ differs from the effective advance $a$ by an amount 
approximately equal to the distortion (in our local time picture $a_{g}=L/c-L/v_{g}$ where $v_{g}$ is the group velocity). 
Finally, the envelope $s(t)$, inverse Fourier transform of $H_{opt}(\omega)E(\omega)$, and the intensity 
profile $\left|s(t)\right|^{2}$ of the transmitted pulse are displayed Fig.\ref{fig3}.

\begin{figure}[htbp]
  \begin{center}
    \includegraphics[angle=-90,width=8.5cm]{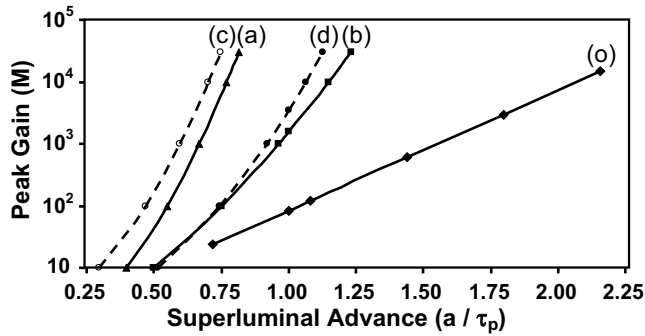}
    \caption {Relation between the peak-gain $M$ and the relative advance
$a/\tau_{p}$ for a given distortion ($D=15\%$). (o) stands for the
optimal system while (a), (b), (c) and (d) respectively relate to the so designated systems (see text). \label{fig4}}
\end{center}
\end{figure}

The efficiency of a superluminal system may be characterised by its
ability to achieve given effects with gains as small as possible.
As above-noticed, the gain of all the optimal systems has the same
asymptotic behaviour ($G\varpropto1/\omega$) and reaches its peak-value
$M$ near $\omega=\pm1$. Consequently $Q$ and $M$ are roughly proportional and can 
indifferently characterise the system gain. The peak-gain $M$, independent of the frequency 
scaling, is retained in the sequel.
This choice facilitates the comparison of the optimal systems with
the most efficient systems used or proposed up to now. 
Since high optical gains exaggerate the problems of instability and noise (amplified
spontaneous emission) and are difficult to achieve with the suitable
frequency-profile \cite{key-13} , we restrict ourselves to systems whose optical
element, responsible for the superluminal effects, is passive. More
specifically, we consider a dilute medium involving (a) an isolated
absorption-line \cite{key-6,key-9,key-10,key-11,key-12}
or (b) a doublet of absorption-line \cite{key-5} and an uniform Bragg-grating
written (c) on a classical optical fibre \cite{key-8} or (d) on
a hollow fibre. Since all these elements are almost transparent outside the low-transmission region (the useful band), 
the peak-gain $M$ is nothing but the gain of the amplifier used to normalise the amplitude of the transmitted pulse. 
The transfer functions are optimised by adjusting the system parameters with a genetic algorithm. A rapid convergence
is obtained by starting the calculations with initial values such that $H(0)=1$ and $a_{g}=a$. For the doublet (b),
the initial value of the line-splitting is chosen such that the $2^{nd}$ order distortion cancels \cite{key-5}. 
Fig.\ref{fig4} shows the results obtained for a reference distortion $D=15$\% and $M$ ranging from $10$ to $3\times10^{4}$. 
No need of a lens to see that the optimal system is much more efficient that the systems (a), (b), (c) and (d) to attain 
large superluminal advances. 
For instance, a peak-gain $M=84$ theoretically suffices to observe an advance $a=\tau_{p}$ with $D=15\%$ (Fig \ref{fig1}) 
but values as large as $1600$, $3400$, $6.4\times10^{6}$ and $4.9\times10^{7}$ would be required with the 
systems (b), (d), (a) and (c) respectively \cite{key-20}. The latter dramatically increase if a lower distortion is required. 
Again for $a=\tau_{p}$ but with $D=7\%$ they raise to $7.9\times10^{4}$,
$2.1\times10^{7}$, $2.3\times10^{14}$ and $4.9\times10^{15}$ while
$M$ only reaches $174$ for the optimal system. By comparison, we stress that achieving experiments with systems 
whose peak amplitude-gain exceeds $10^{4}$ is absolutely unrealistic. 

The situation is much less catastrophic when one examines the superluminal effects which can be attained 
for a fixed peak-gain. Taking $M=1000$ (realisable in a careful experiment) and $D=15\%$ as reference values, 
Fig.\ref{fig4} shows that the relative advance $a/\tau_{p}$ attained with the simplest arrangement 
(medium with an isolated absorption-line) is only $2.4$ times below the theoretical limit ($1.6$) and that the ratio 
falls to $1.7$ by involving a line-doublet. Using non uniform fibre-Bragg-gratings could further reduce this ratio. 
Indeed, at least in principle, these elements allow one to synthesise any transfer function in transmission 
as long as it is minimum-phase \cite{key-21}. This restriction entails that the optimal transfer function (not minimum-phase) 
and thus the upper limit to the advance could be approached but not equalled with these systems. The same remark applies 
to the dispersive media whose transfer function is the exponential of a causal function and is thus also minimum-phase 
\cite{key-5}. Anyway, whatever the system is, superluminal advances exceeding two times the full width at half maximum 
of the pulse intensity-profile are unattainable.

We thank L. Baratchart and F. Seyfert for an helpful discussion on the optimisation procedures and the 
photonic team of PhLAM for useful indications on the fibre-Bragg-gratings. Laboratoire de Physique des Lasers, 
Atomes et Mol\'{e}cules (PhLAM) and Laboratoire Paul Painlev\'{e} are Unit\'{e}s Mixtes de Recherche de l'Universit\'{e} 
de Lille I et du CNRS (UMR 8523 and 8524). CERLA is F\'{e}d\'{e}ration de Recherches du CNRS (FR 2416).


\begin{thebibliography}{10}
\bibitem{key-14}In the pioneering experiments with photonic barriers (see
{[}1, 2{]} and references therein), the barrier lengths were only
few times larger than the optical wavelength. The superluminal advances were then only of the order of the optical period.
\bibitem{key-20}The system (c) suffers from the fact that it is necessary to compensate for the subluminal 
velocity in the host fibre before reaching the superluminal regime.
\end{thebibliography}
\end{document}